\documentclass{ws-procs961x669}            

\begin{document}
\hyphenation{dis-co-ve-red}
\hyphenation{pro-per-ties}

\title{Gamma-ray bursts and their relation to\\
astroparticle physics and cosmology}

\author{J. \v{R}\'{\i}pa}

\address{
Leung Center for Cosmology and Particle Astrophysics, National Taiwan University\\
No.1, Sec.4, Roosevelt Road, Taipei 10617, Taiwan (R.O.C)\\
E-mail: jripa@ntu.edu.tw
}

\begin{abstract}
This article gives an overview of gamma-ray bursts (GRBs) and their relation to astroparticle physics and cosmology. GRBs are the most powerful explosions in the universe that occur roughly once per day and are characterized by flashes of gamma-rays typically lasting from a fraction of a second to thousands of seconds. Even after more than four decades since their discovery they still remain not fully understood. Two types of GRBs are observed: spectrally harder short duration bursts and softer long duration bursts. The long GRBs originate from the collapse of massive stars whereas the preferred model for the short GRBs is coalescence of compact objects such as two neutron stars or a neutron star and a black hole. There were suggestions that GRBs can produce ultra-high energy cosmic rays and neutrinos. Also a certain sub-type of GRBs may serve as a new standard candle that can help constrain and measure the cosmological parameters to much higher redshift than what was possible so far. I will review the recent experimental observations.
\end{abstract}


\bodymatter
\def\figsubcap#1{\par\noindent\centering\footnotesize(#1)}

\section{Gamma-Ray Bursts}\label{sec:grb}
\begin{figure}
\begin{center}
  \parbox{0.48\linewidth}{\includegraphics[width=1.0\linewidth]{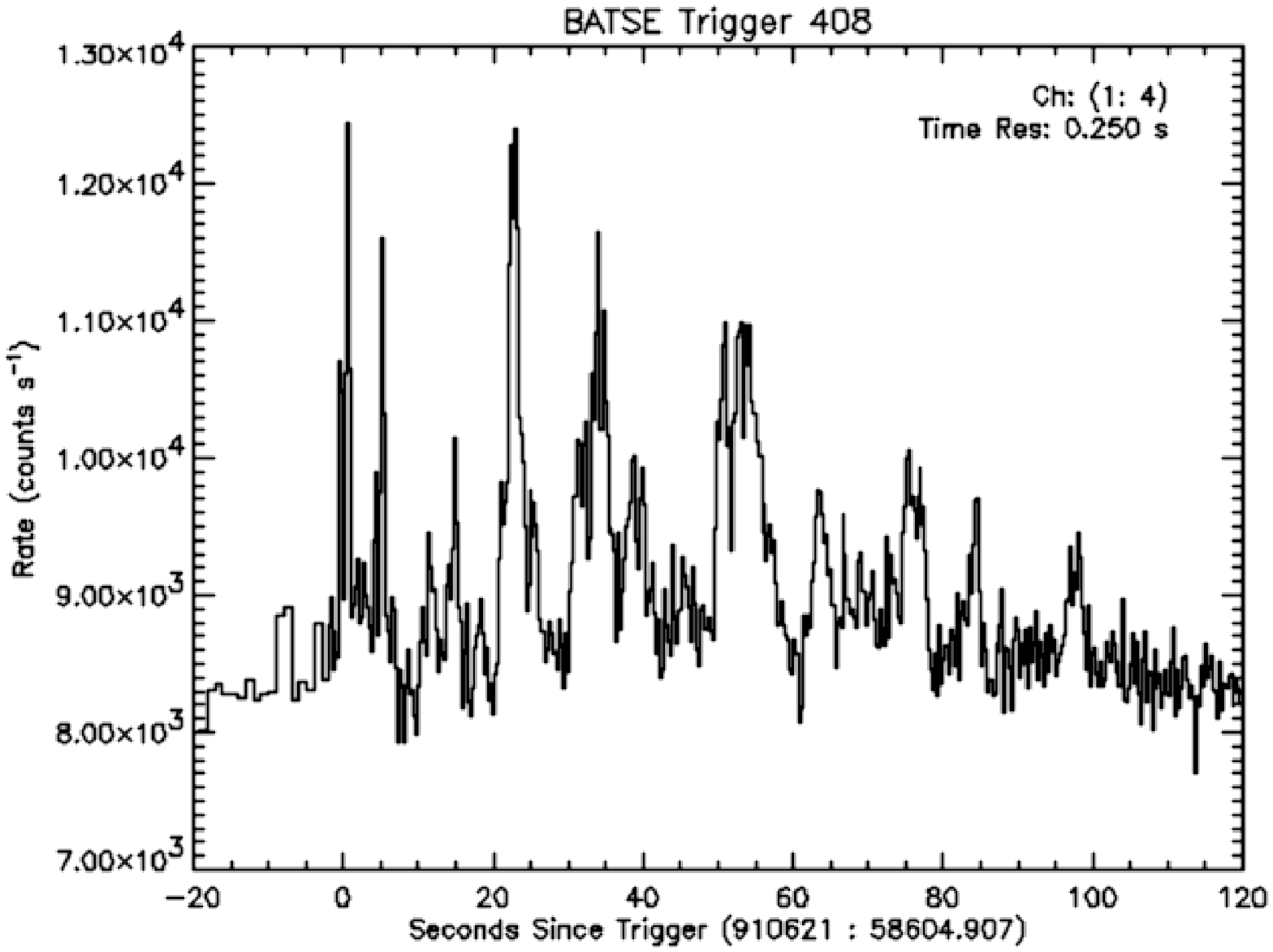}\vspace*{4pt}\figsubcap{a}}
  \hspace*{4pt}
  \parbox{0.48\linewidth}{\includegraphics[width=1.0\linewidth]{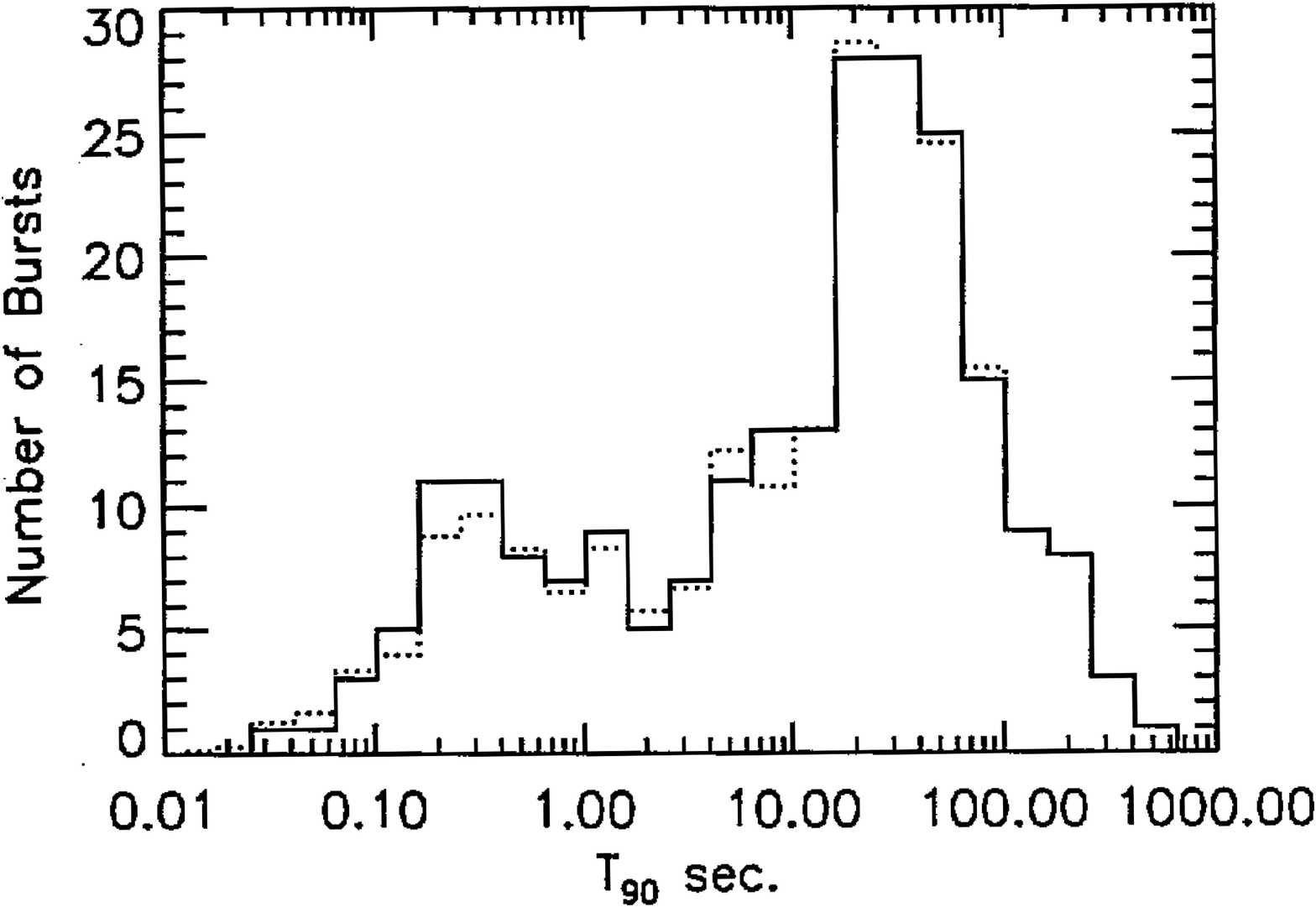}\figsubcap{b}}
  \caption{(a) An example of GRB prompt gamma flux curve detected by BATSE instrument. From ref.\cite{fis94}.
           (b) The distribution of $T_{90}$ durations of GRBs indicating two distinct groups. From ref.\cite{kou93}}
  \label{fig:grb_ltc}
\end{center}
\end{figure}

This section briefly overviews the gamma-ray burst (GRB) phenomena\cite{pir04,mesp06,ved09,kou12,ger13}. GRBs are one of the most extreme explosive events ever observed. They were discovered in late sixties and reported to the scientific community in early seventies\cite{kle73,maz74}. In spite of tremendous effort and numerous observations there are still open questions concerning their detailed physics. They occur roughly once per day and are characterized by flashes of $\gamma$-rays typically lasting from a fraction of a second to thousands of seconds, see \fref{fig:grb_ltc}(a). It has been found that the duration distribution of their $\gamma$-ray prompt emission is bimodial\cite{kou93} which suggested that there were two groups of GRBs, see \fref{fig:grb_ltc}(b). Later, based on more observations, it has been confirmed that these two GRB groups were two distinct astrophysical populations: I. so called long GRBs with prompt $\gamma$-ray emission $\gtrsim 2$\,s that has been identified to be gravitational collapses of massive stars due to their association with type Ic core-collapse supernovae, and II. so called short GRBs with prompt $\gamma$-ray emission $\lesssim 2$\,s that has been suggested to originate in a merger of two compact objects such as NS-NS or NS-BH\cite{ber14}. It was proposed that they lie at cosmological distances\cite{pac86} and they are created in collisions of highly relativistic outflow of the accelerated jetted matter\cite{ree92,ree94}. In many cases it was found that the prompt $\gamma$-ray emission is followed by longer-lasting afterglow in soft X-ray, optical or radio waves\cite{cos97, vpar97} explained as a result of the interaction of the relativistic outflow with the circum-burst medium. Observations found that their redshifts are up to $z=9.4$ with $\langle z \rangle \approx 0.5$ for short GRBs and $\langle z \rangle \approx 2.0$ for long GRBs\cite{ber14} which requires energy release up to an isotropic-equivalent value of $\sim 10^{54}$\,erg.

\section{Relation of GRBs to Cosmology}\label{sec:cosmo}
\begin{figure}
\begin{center}
  \hspace*{4pt}
  \parbox{0.52\linewidth}{\includegraphics[width=1.0\linewidth]{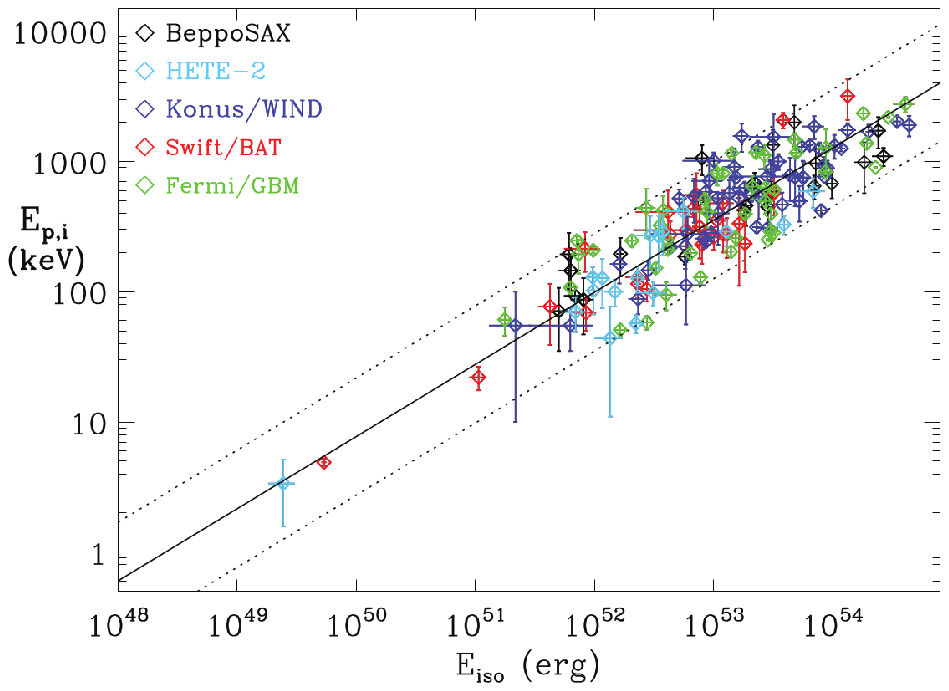}\vspace*{10pt}\figsubcap{a}}
  \hspace*{6pt}
  \parbox{0.41\linewidth}{\includegraphics[width=1.0\linewidth]{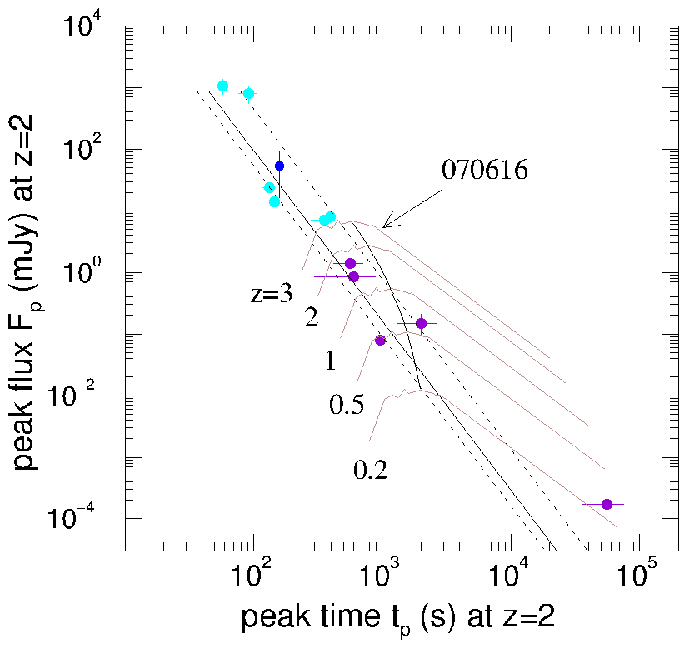}\vspace*{10pt}\figsubcap{b}}
  \caption{(a) The $E_\mathrm{p,i} - E_\mathrm{iso}$ correlation for long GRBs with marked best-fit power-law (black solid line) for different
  instruments. From ref.\cite{ama15}
  (b) The anti-correlation of optical light-curve peak flux (at 2\,eV) for redshift $z=2$ and peak time for six fast rising afterglows (light blue points) and five afterglows slow rising (purple points). The solid straight line shows the best fit. From ref.\cite{panai08}}
  \label{fig:grb_corr}
\end{center}
\end{figure}

\begin{figure}
\begin{center}  
  \parbox{0.42\linewidth}{\hspace*{28pt}\includegraphics[width=1.0\linewidth]{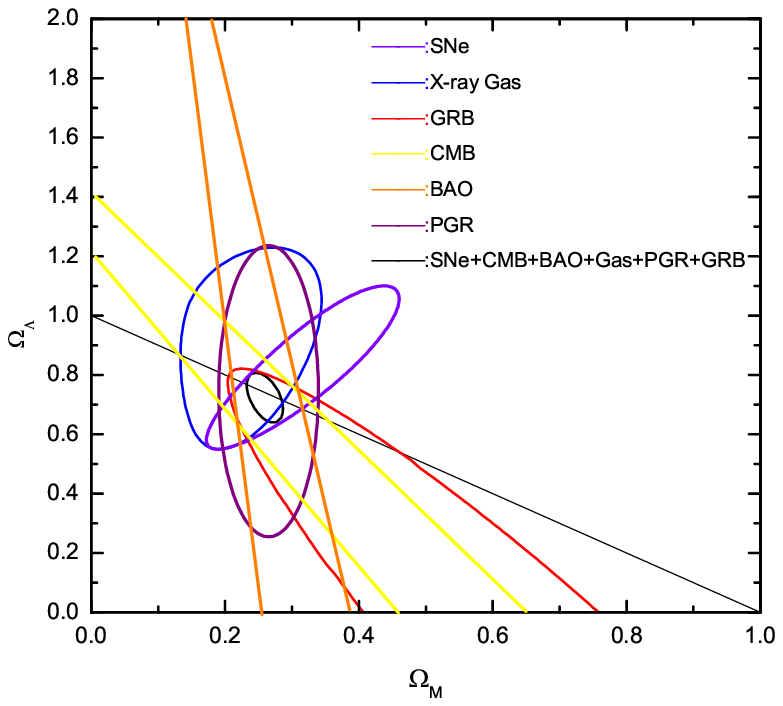}\vspace*{10pt}\figsubcap{a}}
  \parbox{0.55\linewidth}{\hspace*{10pt}\includegraphics[width=1.0\linewidth]{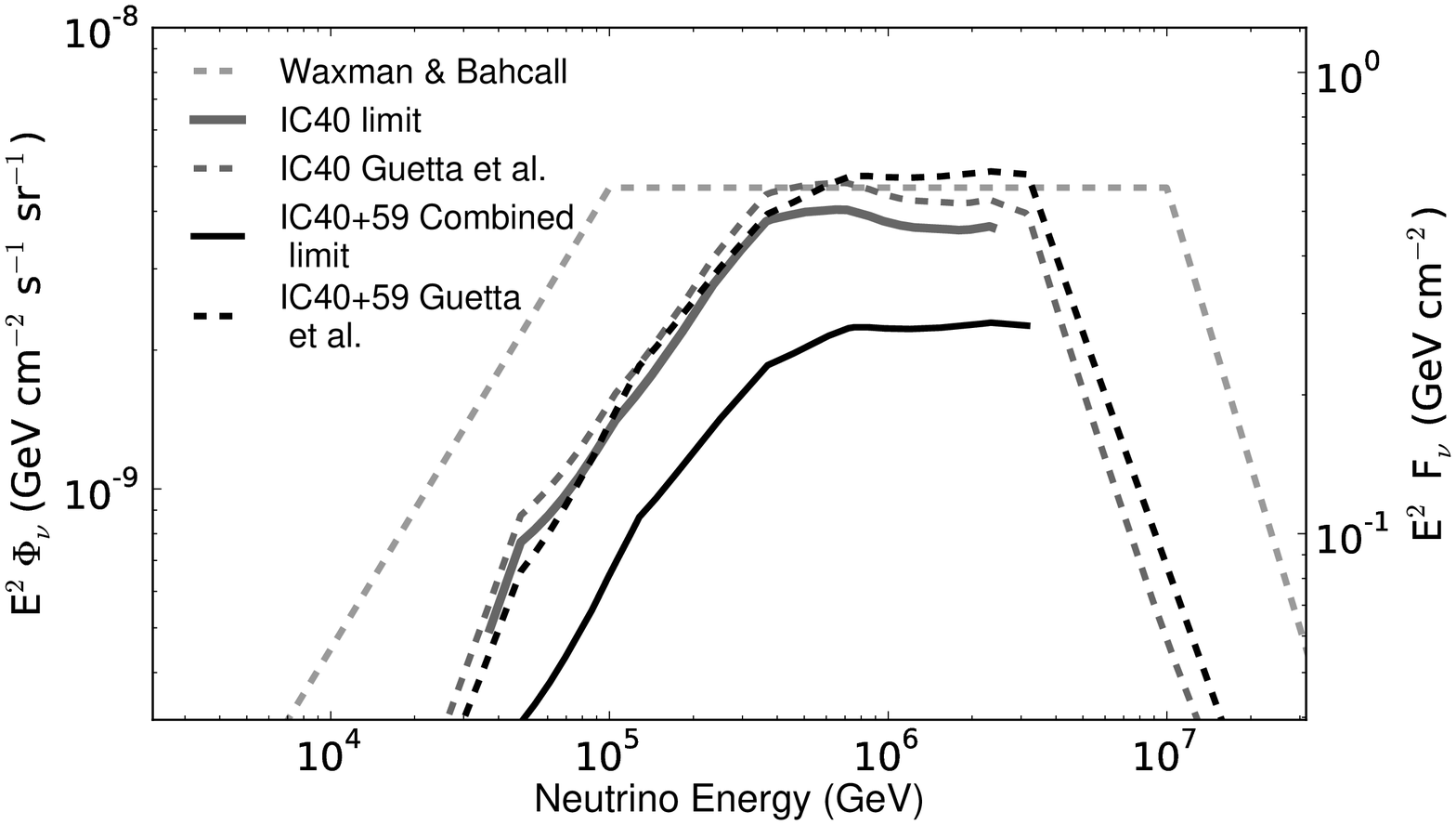}\vspace*{20pt}\figsubcap{b}}
  \caption{(a) Joint $1\sigma$ confidence intervals given by constraints from the datasets of galaxy clusters, GRBs, CMB shift parameter, SNe Ia, BAO, and 2dF Galaxy Redshift Survey. From ref.\cite{wan15,wan07}
  (b) Comparison of predictions of $\nu$ flux from GRBs based on observed $\gamma$-ray spectra (dashed lines) with 90\,\% confidence upper limits obtained from the results of IceCube for 40 and 59 detector strings. From ref.\cite{abb12}}
  \label{fig:grb_cos_par_uhen}
\end{center}
\end{figure}

There were discovered several correlations in the properties of the GRB emission: the anti-correlation between isotropic luminosity $L_\mathrm{iso}$ and spectral lag $\tau_\mathrm{lag}$ of the prompt $\gamma$-ray emission\cite{nor00}; the correlation between $L_\mathrm{iso}$ and light-curve variability $V$\cite{fen00}; the spectral peak energy of the prompt $\gamma$-ray emission $E_\mathrm{peak}$ and isotropic-equivalent released energy $E_\mathrm{iso}$ correlation\cite{ama02}, see \fref{fig:grb_corr}(a); $E_\mathrm{peak}$ and collimated energy $E_\mathrm{\gamma}$ correlation\cite{ghi04}; $E_\mathrm{peak}$ and peak luminosity $L_\mathrm{p}$ correlation\cite{yon04}; correlation between $E_\mathrm{peak}$, $E_\mathrm{iso}$ and rest-frame temporal break in optical afterglow light curve $t_\mathrm{break}$\cite{lia05}; and a correlation between $L_\mathrm{iso}$, $E_\mathrm{peak}$ and prompt 'high-signal' time-scale $T_{0.45}$ (variability)\cite{fir06}. This makes GRBs a potential tool to constrain cosmological parameters\cite{ghi06,wan07,wan15,ama15}, see \fref{fig:grb_cos_par_uhen}(a).

It was also found that the peak flux $F_\mathrm{p}$ (at 2\,eV and $z=2$) might anti-correlate with peak time $t_\mathrm{p}$ in the optical lightcurves\cite{panai08,panai11,panai13}, see \fref{fig:grb_corr}(b). A similar anti-correlation for the peak luminosity in R band of the onset bumps and less steep for the re-brightening bumps in the optical lightcurves was found\cite{lia13}. Unfortunately for many observations the peak time $t_\mathrm{p}$ is unknown and further optical observations at earlier times are needed.

\begin{figure}
\centering
\includegraphics[width=0.4\linewidth]{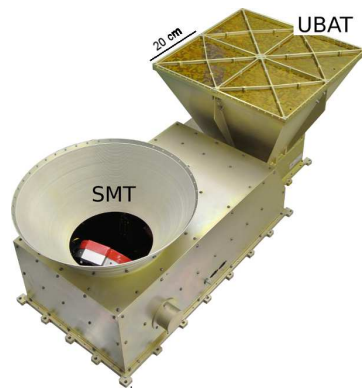}
\caption{The photo shows assembled UFFO-p with its two scientific instruments: UBAT and SMT. From ref.\cite{rip15}}
\label{fig:uffo}
\end{figure}

More early optical observations with fine sapling of the light curves around the peak times below one minute could be provided by the Ultra-Fast Flash Observatory pathfinder (UFFO-p)\cite{che11,par13,nam13}. It is a novel spaceborn instrument dedicated to detect GRBs and rapidly follow their optical/ultraviolet counterparts to provide prompt optical and early afterglow measurements. It consists of two scientific instruments, see \fref{fig:uffo}. A GRB location is determined in a few seconds by the first instrument called UFFO Burst Alert \& Trigger telescope (UBAT) \cite{jun11,lee13,cha14,rip15} which employs the coded mask imaging technique \cite{con13} and the detector comprising of Yttrium Oxyorthosilicate scintillating crystals and multi-anode photomultiplier tubes. It has the energy range of $\approx 10-150$\,keV with half-coded field of view (HCFOV) $70.4^{\circ}\times70.4^{\circ}$ and angular resolution $\le10'$. The second instrument is the Slewing Mirror Telescope (SMT) \cite{jeo13a,jeo13b,jeo13c,kim13} with the field of view of $17'\times17'$ and UV range of $200-650$\,nm. It consists of a Ritchey-Chr\'{e}tien telescope with an Intensified Charge-Coupled Device in its focal plane. In front of the telescope there is placed a fast plane slewing mirror which allows to redirect the optical path and start observation within $\approx 1$\,s since it receives the target direction. SMT slewing mirror provides approximately the same sky coverage as UBAT's HCFOV. UBAT and SMT have been assembled and integrated with the control electronics on UFFO-p which is planned to be launched on the Lomonosov Moscow State University satellite \cite{panas11}.

\section{Relation of GRBs to Astroparticle Physics}\label{sec:astropart}

GRBs were proposed to be sites for accelerating ultra-high energy cosmic rays and sources of very high energy neutrinos up to $10^{17}\sim10^{19}$\,eV \cite{wax97,wax00,mesp01,wax03,mesp15}. However, the recent results from the IceCube detector\cite{gue15} suggests that the efficiency of neutrino production may be much lower than predicted\cite{abb12}. 4-years data of IceCube put constraints on the prompt neutrino flux from GRBs, see \fref{fig:grb_cos_par_uhen}(b). A single low-significance neutrino, compatible with the atmospheric neutrino background, was found in coincidence with one of the 506 observed GRBs\cite{aar15}.

\end{document}